\documentclass[
    twocolumn,
    aps,
    prl,
    superscriptaddress,
    longbibliography,
    10pt
]{revtex4-2}

\usepackage{graphicx,color,xcolor,physics}
\usepackage[
    colorlinks=true,
    urlcolor=blue,
    linkcolor=blue,
    citecolor=blue,
    breaklinks,
]{hyperref}

 \usepackage{amsmath}
 \usepackage{amsfonts}
 \usepackage{amssymb}
 \usepackage{braket}
 \usepackage{graphicx}
  \usepackage{hyperref}

\begin{document}

\newcommand{\tm}[1]{{\color[rgb]{0.6,0.0,0.0}#1}}\textbf{}
\newcommand{\fd}[1] {{\color[rgb]{0,0.4,0.65}#1}}
\newcommand{\Maedeh}[1]{\begingroup\color{violet}#1\endgroup}
\newcommand{\alex}[1]{{\color[rgb]{0.0,0.6,0.0}#1}}

\title{
Freeness Reined in by a Single Qubit
}

\author{Alexander Altland}
\affiliation{Institut f\"ur Theoretische Physik, Universit\"at zu K\"oln, Z\"ulpicher Straße 77, 50937 K\"oln, Germany}

\author{Francisco Divi}
\affiliation{Perimeter Institute for Theoretical Physics, Waterloo, Ontario N2L 2Y5, Canada}
\affiliation{ICTP South American Institute for Fundamental Research, Instituto de F\'{i}sica Te\'{o}rica,\\
Univ. Estadual Paulista, Rua Dr. Bento Teobaldo Ferraz 271, 01140-070, S\~{a}o Paulo, Brazil} 

\author{Tobias Micklitz}
\affiliation{Centro Brasileiro de Pesquisas Físicas, Rua Xavier Sigaud 150, 22290-180, Rio de Janeiro, Brazil}

\author{Maedeh Rezaei}
\affiliation{Institut f\"ur Theoretische Physik, Universit\"at zu K\"oln, Z\"ulpicher Straße 77, 50937 K\"oln, Germany}

\date{\today}

\begin{abstract}
Free probability provides a framework for describing correlations between
non-commuting  observables in complex quantum systems whose
Hilbert-space states follow maximum-entropy distributions. We examine
the robustness of this framework under a minimal deviation from freeness:
the coupling of a single ancilla qubit to a Haar-distributed quantum
circuit of dimension $D_0 \gg 1$. We find that, even in this setting, the
correlation functions predicted by free probability theory receive
corrections of order $\mathcal{O}(1)$. These modifications persist at
long times, when the dynamics of the coupled system is already ergodic.
We trace their origin to non-uniformly distributed stationary quantum states, which we characterize analytically and confirm numerically.
\end{abstract}

\maketitle

\emph{Free probability} has emerged as a powerful new framework describing
statistical correlations beyond Gaussianity in complex quantum systems.
Originally developed in the context of random matrices and operator
algebras~\cite{Voiculescu1991,Speicher1997,CollinsSniady2006,Guionnet2014}, it
extends the notion of classical independence to non-commuting quantum
observables: in the  
limit of large Hilbert-space dimension $D\to \infty$,  
independent random operators become free in the sense of mutually orthogonal
eigenbases and factorizing operator cumulants. 

While the mathematics of free probability of maximally random states is well
 under control, the applied perspective of the concept lies in the description
 of  systems departing from full ergodicity, either transiently before
 approaching a late time ergodic limit, or permanently. Use cases, include the
 statistics of few-body observables in the context of the eigenstate
 thermalization hypothesis
(ETH)~\cite{deutsch1991quantum,srednicki1994chaos,rigol2008thermalization,pappalardi2022eigenstate,Cotler2017,d2016quantum, kim2014testing} or, the identification
 of approximately  free-probabilistic structures in the dynamics of random
 quantum circuits~\cite{fava2025designs, fritzsch2025free}.
 
 In these applications, one is generically met with structures whose eigenstate
 distributions  may be random, but generically not fully uniformly distributed
 in Hilbert space. The stability of  free probability to such deviations from maximum entropy limits is only beginning to be
 investigated~\cite{ChalkerPhysRevX.14.031029,yoshimura2025operator,
 bouverot2025random,BernardPhysRevX.13.011045}: will mild deviations from an
 ergodic limit --- as quantified, e.g., by differences from random matrix
 statistics --- reflect in equally small changes in the statistical cumulants
 describing freeness, or should we expect more drastic consequences? 

In this Letter, we systematically address this question for the `smallest'
conceivable deviation from a free limit, a single ancilla qubit, or spin,
coupled to a fully random environment of dimension $D_0\gg 1$. For a golden rule 
coupling strength $\gamma$ exceeding the average level spacing, $\Delta_0$, of
the latter, this system will approach an ergodic long-time limit at a timescale
set by  $\sim \gamma^{-1}$. Our main finding is
that, perhaps unexpectedly, the presence of the single degree of freedom not
only remains visible in the limit of large $D_0$, but actually leads to
$\mathcal{O}(1)$ changes in statistical correlators including time scales where
the dynamics has become ergodic. We will analytically and numerically describe
these deviations, and provide physical interpretations of their significance.

\begin{figure}[t!]
    \centering
    \vspace{-.0cm}
    \includegraphics[width=0.6\linewidth]{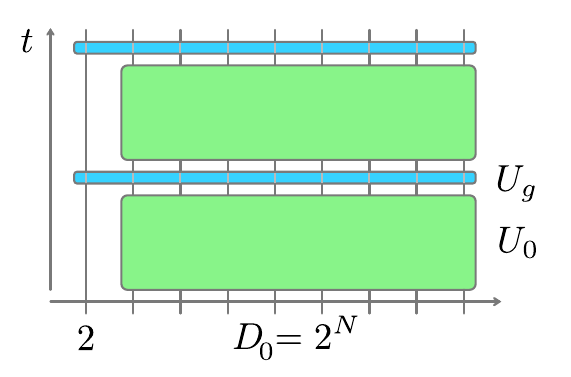}
    \vspace{-.4cm}
    \caption{A single ancilla qubit coupled to a $D_0=2^N$-dimensional $N$-qubit
    circuit subject to Haar unitary time evolution. The unitary coupling
    operator $U_g$ realizes an interaction of tunable strength, $g$,
    leading to generally non-uniform quantum state distributions.}    
\label{fig:figure1}
\end{figure}

{\it Model:---}Our reference model contains an $N$-qubit quantum circuit
subject to  discrete time evolution governed by a
Haar-distributed~\cite{plodzien2025lecture} $U_0=\{U_{\mu \nu}\}$, weakly
coupled to a single qubit via a unitary  $U_g$. Describing the latter as
  \begin{align}
    \label{eq:circuit_model}
    U_g &= e^{-i \left(
    W\otimes\sigma^+ + W^\dagger\otimes\sigma^- \right)},
\end{align}
in terms of a Gaussian distributed operator with variance $ \overline{\tr(W W^\dagger) } =D_0g^2/2$
coupled to the qubit-spin raising and lowering operators $\sigma^\pm$, we
consider the Floquet evolution $U =  U_g (U_0\otimes \openone_2)$, Fig.~\ref{fig:figure1}. 

Now, consider two operators $A$ and $B$ defined in the Hilbert space
$\mathbb{C}^{D_0}\otimes \mathbb{C}^2$ of the composite system, and assumed to be traceless $\tr_{D_0}(X)=\tr_2(X)=0$, $X=A,B$ individually in
the factor spaces for simplicity. Free probability reduces the Haar averaged
expectation values $\overline{\left\langle (A B_t)^k\right\rangle }$ of traces
$\left\langle \dots \right\rangle\equiv D^{-1}\mathrm{tr}(\dots)$ over products
of $A$ and the time evolved $B_t \equiv U^t B U^{\dagger t}$, to sums of
cumulants of the two bare operators, weighted by eigenphase correlation
functions of $U^t$ (see the Supplemental Material for a short review of the
concept).  The simplest of these relations assumes the form
\begin{align}
  \label{eq:free_2pt_traceless}
\overline{\langle A B_t\rangle}
&=
\,K(t)\langle AB\rangle, \qquad
    K(t)
    = \overline{|\langle U^t\rangle|^2}, 
\end{align}
where the form factor, $K(t)$, probes the second moment of spectral correlations
of $U$ in the time-Fourier transform domain \footnote{Inspection of this
relation for $t=1$ shows that in discrete time, the right-hand side contains
$K(t)-D^{-2}$. However, assuming $t\gg1 $ we ignore this subtlety.}. From the perspective
of many body physics, the left-hand side, $\sim \mathrm{tr}(A B_t) $, is a
dynamical  correlation function, while the right-hand side $\sim
|\mathrm{tr}(U_t)|^2 $ probes spectral correlations, i.e. quantities of
thermodynamic significance. Free probability makes the remarkable statement that
these quantities can be equal under  conditions of maximal
entropy. To explore the scope of their equivalence in the present setting, we
now analyze both sides of Eq.~\eqref{eq:free_2pt_traceless}  separately. 

\begin{figure}[t!]
    \centering
    \vspace{-.0cm}
    \includegraphics[width=0.99\linewidth]{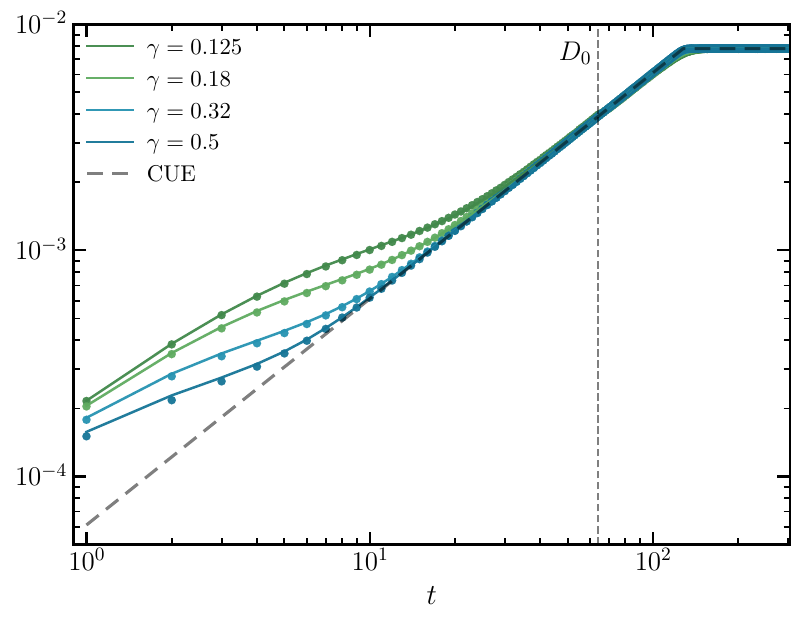}
    \vspace{-.8cm}
    \caption{Numerically computed spectral form factor $K(t)$ (dots),
    for $D_0=64$ and coupling strengths $\gamma$, averaged over $10^6$ realizations, compared to the prediction Eq.~\eqref{eq:SFF_ancilla}  extended to the full time domain by
    Eq.~(16) of the Supplemental Material (solid lines).}
\label{fig:figure2}
\end{figure}

\textit{Form factor:---}Beginning with the form factor, in ergodic random
matrix theory, $K(t)$ shows the notorious ramp-plateau profile \cite{haake2018}: linear increase
in time $K(t)=t/D^2$, which at the Heisenberg time $t=D$ abruptly ends in a
stationary plateau $K(t)=1/D$. More generally, deviations from linearity signal
ergodicity violations \cite{plodzien2025lecture,haake2018}, at times  $t
\lesssim t_\mathrm{Th} $ defining an effective Thouless time scale.

\begin{figure}[h!]
    \centering
    \vspace{-.0cm}
    \includegraphics[width=0.99\linewidth]{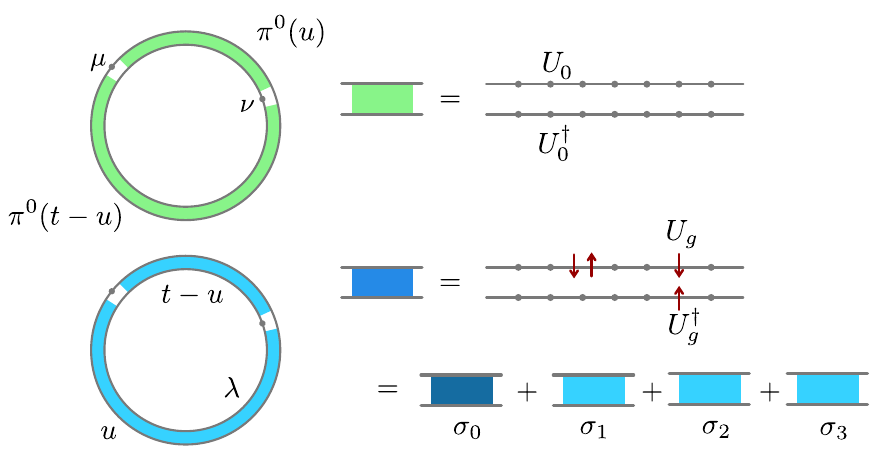}
    \vspace{-.8cm}
    \caption{
    Left: the spectral form factor schematically represented in terms of two
    interfering closed loop quantum amplitudes of discrete propagation time $t$.
    The gray shaded double (top) lines represent the sequential application of the
    random operator $U_0\otimes U_0^\dagger$  in a non-fluctuating singlet channel.
    The inclusion of the ancilla qubit  (right center) stabilizes $4=1+3$ modes
    operating in conserving spin singlet and triplet channels, respectively.
  }    
\label{fig:figure3}
\end{figure}

Fig.~\ref{fig:figure2} shows the form factor of our system for various values of
the golden rule coupling parameter $\gamma = g^2/2$~\footnote{The evaluation of the coupling parameter via Fermi's golden rule yields $\gamma =\frac{1}{D_0}
\sum_{\mu} \Gamma_{\mu} \sim 
%\frac{1}{D_0} \sum_{\mu} \sum_{\nu}
%\overline{|\expval {\nu , 1|W\sigma ^+|\mu , %0}}|^2=
\frac{1}{D_0}
\overline{\tr(WW^{\dagger})}=  \frac{g^2}{2}$, where we have performed an average over the initial states $\mu$ and a sum over the final states $\nu$, together with a disorder average over the interaction matrix elements, for which $\overline{|W_{\mu\nu}|^2} =
\frac{g^2}{2D_0}$}, where we used that the mean level spacing of the bulk
system operator $U$ is given by $\Delta_0=2\pi/D_0$. For times exceeding $\sim
\gamma^{-1}$, the form factor relaxes to an ergodic profile with the Heisenberg
time $t=D=2D_0$ reflecting the dimension of the full Hilbert space, including
the qubit. Interestingly, these curves can be obtained analytically with little
effort from an intuitive semiclassical construction:

Considering the environment in the absence of the ancilla first, the linear time
dependence of
 the form factor $K(t)=D_0^{-2} %\big\langle 
 \sum_\mu \sum_\nu
 \overline{ \braket{\mu|U_0^t|\mu} 
 \braket{\nu |  U_0^{\dagger t}|\nu}}$
 %\big\rangle  
 results from the  interference of  quantum amplitudes 
 $\ket{\mu}\stackrel{t}{\rightarrow }\ket{\mu}$ and their complex conjugates propagating
 along closed loops of step length $t$ in Hilbert space, Fig.~\ref{fig:figure3}. Loop-pairs
 surviving the average over rapid phase fluctuations must be piece-wise
identical, as indicated in the top left panel of Fig.~\ref{fig:figure3}.
Denoting pair mode segments of discrete time duration $0\le u\le t$ by $\pi_0(u)$,
causality, ergodicity, and probability conservation require
$\pi_0(u)=\Theta(u)/D_0$, where $\Theta$ is the Heaviside step function. Summation over the
intermediate time, then yields the form factor as a time-convolution $K(t)=
(\pi_0 \ast \pi_0)(t)\equiv  \sum_{u=1}^t
\pi_0(t-u)\pi_0(u)=t/D_0^2 $.

 Upon including the ancilla, $U_0\to U_g\left(U_0\otimes \openone_2\right)\equiv
 U$, the double line `particle-hole propagators' acquire a spin structure
 $\pi_0\to \{\pi_{a,b}\}$, $a,b\in \{0,1\}$ reflecting the coupling of our two
 quantum amplitudes to a binary degree of freedom. These $2\times 2$ matrices
 can be organized in Pauli channels, $\pi = \sum_{\lambda=0}^3\pi_\lambda
 \sigma_\lambda$, where  the singlet mode $\pi_0$ is isotropic in spin-space and
 the
 three triplet modes $\pi_{1,2,3}$ correspond to modes of conserved total spin
 $1$.

 The quantitative computation of these modes amounts to a perturbation
 series summation,  a task we solve by path integral methods, detailed in the
 Supplemental Material. This computation demonstrates the conservation of the
 total mode spin $S=0$ for $\pi_0$ and $S=1$ for $\pi_i$, a fact owed to the structure of  the
  operator $\sim(\sigma^+\otimes
 \sigma^-+\mathrm{h.c.})$ averaged over realizations of the coupling matrix $W$.
Quantitatively, we obtain the time dependence 
\begin{align}
  \label{eq:quantum_modes}
\pi_\lambda(u)
=\frac{\theta(u)}{D}e^{-\gamma_\lambda u},
\qquad 
\gamma_0=0, \quad \gamma_{1,2} = \gamma, \quad \gamma_3=2\gamma,
\end{align}
where the stationary spin-singlet $\pi_0$ assumes the role
of an ergodic mode in a Hilbert space of doubled dimension $D=2D_0$, while the
spin-triplet modes $\pi_i(u)$  exponentially decay and the degeneracy
$\gamma_1=\gamma_2$ reflects the $3$-axis rotational invariance of the coupling.
Spin conservation further implies additivity in the form factor, which now reads
\begin{align}
  \label{eq:SFF_ancilla}
    K(t)
    &= 
    \sum_{\mu=0}^3 (\pi_\lambda*\pi_\lambda)(t)=\frac{t}{D^2}\left( 1+2e ^{-\gamma t}+ e^{-2\gamma t} \right).
\end{align}

For values  $\gamma \gtrsim \Delta$ describable by the golden rule --- for smaller
 values, the ancilla decouples in a manner we have not investigated in detail
 --- and perturbatively accessible time scales $t<D$, Eq.~\eqref{eq:SFF_ancilla} is in  parameter-free
 agreement with our numerical results. Within our
 path integral formalism, larger (``plateau'') time scales are described by
 non-semiclassical saddle points~\cite{Andreev1997}, whose inclusion leads to
 the generalized formula Eq.~(16) of the Supplemental material. This extension
 (plotted as solid lines) agrees with the numerics over the entire time domain.

\begin{figure}[h!]
    \centering
    \vspace{.1cm}
    \includegraphics[width=4.0cm]{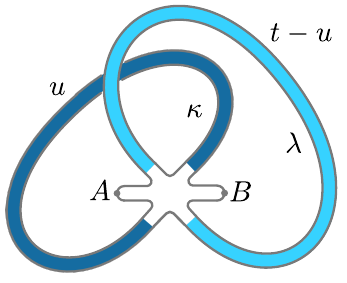}
    %\vspace{-.2cm}
    \caption{Two-loop diagram providing the dominant contribution to the
    correlation function $\overline{\langle A B_t \rangle} $. Note the doubly self
    retracing loop structure, which in the spinless case makes the expression
    proportional to $\left\langle AB \right\rangle $. Here, the
    coupling of the mode-spin channels to the observables leads to a nested
    structure $\sim \langle B\sigma_\lambda \sigma_\kappa A
    \sigma_\lambda
    \sigma_\kappa \rangle  \pi_\lambda * \pi_\kappa$} .   
\label{fig:figure4}
\end{figure}

{\it Correlation function:---}Turning to  the left-hand side of Eq.~\eqref{eq:free_2pt_traceless}, the
application of the  mode interference rationale previously applied to the
form factor reveals the structure shown in Fig.~\ref{fig:figure4} as the
dominant contribution to the correlation function of individually traceless
 observables, $\left\langle A \right\rangle = \left\langle B \right\rangle=0  $.
From the perspective of ergodic slow modes, this expression resembles
$\left\langle AB \right\rangle \times $ (form factor): the product of two modes of individual
duration $u$ and $t-u$. This equivalence defines the semiclassical
interpretation of the  identity
Eq.~\eqref{eq:free_2pt_traceless}.  

However, the equivalence between the two
sides of the equations ends when we consider structures beyond the ergodic limit:
The pair modes contributing to the correlation function revisit the same point
in configuration space \emph{twice}, while the form factor is described by a
single self retracing loop. In the parlance of effective field theory, this is
the difference between a one- and a two-loop process, which can be major in
settings outside the maximum entropy limit. Additionally, the spin structure of our modes is
contracted against that of the observables $A$ and $B$, in a manner different
from that in the form factor. 

That these differences may lead to significant consequences 
is substantiated by the counting of Hilbert space
dimensions. In a many-body context, the tensorial coupling of just a single qubit
doubles the Hilbert space dimension $D_0 \to 2 D_0$. The resulting injection of a large number of
correlated matrix elements into a previously homogeneously distributed
background, combined with the different loop order, affects the correlation
function in a manner that we now quantify.

Consider a representation of the  
correlation function as   
\begin{align}
  \label{eq:minimal_model_2pt}
  \overline{\langle A B_t \rangle}
  &=
  K(t)\langle AB\rangle
  +\Delta(t),
\end{align}
where the first term  extends the free-probability result
Eq.~\eqref{eq:free_2pt_traceless} to include the triplet modes in the form factor
Eq.~\eqref{eq:SFF_ancilla}. For the second term, we find 
\begin{align}
  \Delta(t)
  &= 2 \sum_{(ijk), \mathrm{cycl.} }
  F_{ijk}(t)\langle A\sigma_k B\sigma_k \rangle ,
  \label{eq:DeltaAB}
\end{align}
where $F_{ijk}= \pi_0 * \pi_k - \pi_i * \pi_j $, and  `$\mathrm{cycl.}$'
denotes a summation over cyclically permuted indices, $(ijk)=(123),(231),(312)$.
 The structure of Eqs.~\eqref{eq:minimal_model_2pt}
and~\eqref{eq:DeltaAB} follows from that of the diagram shown in
Fig.~\ref{fig:figure4}. Summing over the different configurations of
individually spin-conserving modes, the central vertex assumes the form $\sim
\tr(A \sigma_\lambda \sigma_\kappa B \sigma_\lambda \sigma_\kappa) \times
\pi_\lambda(u)\pi_\kappa(t-u)$, where the Pauli matrix pairs $\sigma_\lambda $ and
$\sigma_\kappa$ decorate the endpoints of the mode shown in light and dark blue,
respectively. The identity of these Paulis determines the time dependence,
$\pi_\lambda(u)$ and $\pi_\kappa(t-u)$, and their commutation relations the algebraic
structure of the two equations, where the term $K(t) \left\langle A B
\right\rangle $ is the contribution of identical modes $\sigma_\lambda \sigma_\lambda
=1$. 

The remaining time convolution integrals are elementary, and we find
\begin{align*}
 F_{312}(t)=F_{231}(t)&= \frac{( 1 - e^{-\gamma t})^2}{\gamma D^2}  \stackrel{t\to \infty}\longrightarrow  \frac{1}{\gamma D^2},\\
F_{123}(t)&= \frac{ 1 - e^{-2\gamma t}}{2\gamma D^2}  - \frac{t  e^{-\gamma t}}{D^2} \stackrel{t\to \infty}\longrightarrow  \frac{1}{2\gamma D^2}. 
\end{align*}
For intermediate time scales $t\sim \gamma^{-1}$, the functions $F_{ijk}$ add to
a correction $\Delta(t)\sim 1/(D^2 \gamma)\sim t/D^2 \sim K(t)$ of the same
parametric order as the form factor itself: the free probability identity is
violated at an $\mathcal{O}(1)$ level. In addition, the mismatch $\Delta(t)$
does not decay, it saturates at a base value $\sim 1/(D^2\gamma)$, implying a
deviation from Eq.~\eqref{eq:free_2pt_traceless}, including for time scales
$t\gg \gamma^{-1}$, where the  dynamics has become ergodic. 

\begin{figure}[t!]
    \centering
    \includegraphics[width=0.99\linewidth]{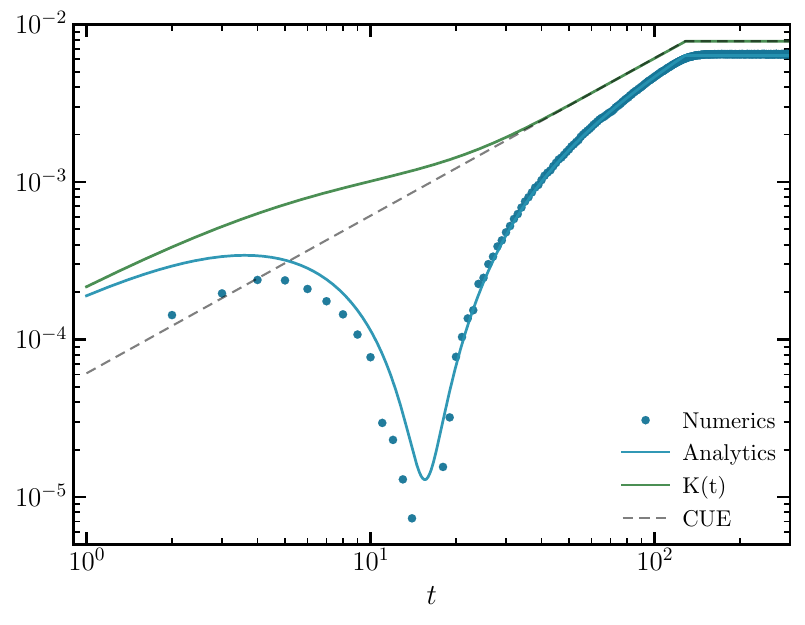}
    \vspace{-0.6cm}
    \caption{Numerical validation of two-point function Eq.~\eqref{eq:minimal_model_2pt}, here 
    for $A=B= \sigma_3\otimes \sigma_3 \otimes 1_{D_0/2}$ 
    computed numerically for $\gamma = 0.125$, $D_0 = 64$ and an ensemble-average over 
    $10^6$ realizations. The green line shows  
    the spectral form factor, 
    and the dashed line is the free probability prediction Eq.~\eqref{eq:free_2pt_traceless}. }
    \label{fig:figure5}
\end{figure}

Fig.~\ref{fig:figure5} shows that these predictions are in good agreement with
numerical observation. In particular,  the constant mismatch relative to the
form factor is confirmed by the numerics, including for time scales past the
Heisenberg time $t> D$. At first sight,  the deviation from statistical
freeness, extending deep into the ergodic regime, may be startling. To
understand that it is consistent with ergodicity, we need to carefully
distinguish between ergodicity in time evolution and the stronger condition of
uniform state distribution and statistical freeness. To this end, let us rewrite
the correlation function as $\left\langle A B_t  \right\rangle=
\frac{1}{D}\tr_2((A\otimes B) \overline{(U\otimes U^\dagger)} S  )  $,  where
$U=((U_0\otimes \openone_2)U_g)^t$ is the full time evolution operator,
$\tr_2$ is the trace over two tensor copies of our Hilbert space, and
$S\ket{\alpha}\otimes \ket{\beta}=\ket{\beta}\otimes \ket{\alpha} $ with
$\ket{\alpha}=\ket{\lambda,a}$ the swap operator. Reformulated in this way, we
may $1/D$-expand the configuration-averaged time evolution as
\begin{align}
  \label{eq:TensorTimeEvolution}
  \overline{U\otimes U^\dagger}= \pi_0 S + (\pi_0\ast\pi_0)\openone  + (\pi_0\ast \pi_i) (\sigma_i \otimes
\sigma_i)+\dots,
\end{align} 
in the unitary invariants of the environment space. Here, the coefficients weighing
individual Pauli channels can be obtained by matching to our previous mode
analysis, or by direct series summation. Specifically, the first term leads to a
contribution $\tr_2((A\otimes B)S^2)=\tr(A)\tr(B)$, vanishing for traceless observables. The
second term yields the long-lived contribution to the form factor $\pi_0 \ast
\pi_0 \sim t$. At the next leading order, we encounter convolutions of the ergodic
mode into decaying modes $\pi_0 \ast \pi_i \propto \gamma^{-1}$, which
contribute to the correlation function, including in the stationary long-time
limit.

By contrast, the substitution of Eq.~\eqref{eq:free_2pt_traceless} into the form factor
$K(t)=\tr_2(\overline{U\otimes U^\dagger})$ gets us back to our previous conclusion
that mode coupling terms such as $\pi_0 \ast \pi_1$ cancel out due to the
vanishing trace over uncompensated Pauli operators; Unlike the correlation
function, the form
factor  reduces to the ergodic contribution $\pi_0\ast \pi_0$ at time scales $\gtrsim
\gamma^{-1}$. The upshot of this consideration is that the decay of all  slow
modes --- dynamical ergodicity as witnessed by the form factor ---  does not suffice to
establish statistical freeness: the latter also requires a uniform state distribution
  over the full Hilbert space, which is a stronger
condition not holding in the
present setting.

One may then ask under what conditions the system comprising the ancilla \emph{will}
become equivalent to a single maximum entropy random matrix ensemble of
dimension $2D_0$. Our previous analysis indicates that this is the case for
coupling strength $\gamma \gtrsim 1$, of the order of the bandwidth of the environment operator. In
this limit, the mismatch function $\Delta(t)$ decays almost instantly to a base
value $\sim 1/D^2$ smaller by a factor $\sim D$ than the form factor $K(t)$ for
characteristic ramp times $t\sim D$. Another way to arrive at the same
conclusion is to observe that $\gamma \sim |W|^2 D$ sets the Born
hybridization energy scale of the typical ensemble states due to the coupling to
the ancilla. If this becomes of the order of the quasienergy bandwidth $2\pi$, the coupling leads to complete state scrambling, and we should expect statistical freeness.

{\it Summary and Discussion:---} We have shown how the coupling of a single
degree of freedom to a $D$-dimensional maximum-entropy environment may induce
$\mathcal{O}(1)$ changes in equalities between correlation functions derived
under the assumption of statistical freeness. These changes were controlled by
the ancilla-environment coupling rate, $\gamma$, with stationary long-time
limits beyond the system's ergodic time $\sim \gamma^{-1}$. While we derived our
results for the lowest nontrivial correlation function, $\overline{\left\langle
(A B_t)^k\right\rangle }$ for $k=1$, we expect the underlying principle ---
non-uniform state distributions in the qubit--environment space --- to affect
correlations of higher order, $k>1$, in a similar manner. In this sense, free
probability provides a statistical framework responding sensitively to
departures from the maximum-entropy limit of uniform quantum state
distributions, which is a stronger criterion than dynamical ergodicity. (For a
conceptually related study, see Ref.~\cite{logaric2024quantum}.)  
It will be interesting to explore applied consequences of these findings, 
for example, in the analysis of correlation matrices, entering the quantitative 
formulation of the eigenstate thermalization hypothesis. We finally note that
the main protagonist of our story, the two-loop quantum interferece process of
Fig.~\ref{fig:figure4}, has a history in the physics of random quantum
transport. In applications where $A,B$ are current operators through a random
scattering medium,  it features as an antagonist of the coherent backscattering
peak (`weak localization'),  describing a quantum-enhanced probability to
stay in a \emph{forward} scattering channel
\cite{micklitz2014a,Karpiuk2012}. This is in line with the
interpration of the diagram Fig.~\ref{fig:figure4} as the leading contribution
supporting the mutual coherence of two operators, now in
a many-body space.

\emph{Acknowledgments.---} Special thanks go to S.~Pappalardi and D.~Rosa for
valuable input during the initial stages of this work. T.~M.~acknowledges financial support by Brazilian
agencies CNPq and FAPERJ. A.~A. acknowledges partial support from the Deutsche
Forschungsgemeinschaft (DFG) under Germany’s Excellence Strategy Cluster of
Excellence Matter and Light for Quantum Computing (ML4Q) EXC 2004/1 390534769
and within the CRC network TR 183 (project grant 277101999) as part of projects
A03. F.~D.~would like to thank the Perimeter/ICTP-SAIFR/IFT-UNESP fellowship
program and CAPES for financial support, and acknowledges the
hospitality of the Institute for Theoretical Physics at the University of
Cologne, where part of this work was carried out.

\bibliography{ancilla_spin_refs.bib}

\end{document}

\title{Supplemental Material to Free Probability Reigned in by a Single Qubit
}

\author{Alexander Altland}
\affiliation{Institut f\"ur Theoretische Physik, Universit\"at zu K\"oln, Z\"ulpicher Straße 77, 50937 K\"oln, Germany}

\author{Francisco Divi}
\affiliation{Perimeter Institute for Theoretical Physics, Waterloo, Ontario N2L 2Y5, Canada}
\affiliation{ICTP South American Institute for Fundamental Research, Instituto de F\'{i}sica Te\'{o}rica,\\
Univ. Estadual Paulista, Rua Dr. Bento Teobaldo Ferraz 271, 01140-070, S\~{a}o Paulo, Brazil} 

\author{Tobias Micklitz}
\affiliation{Centro Brasileiro de Pesquisas Físicas, Rua Xavier Sigaud 150, 22290-180, Rio de Janeiro, Brazil}

\author{Maedeh Rezaei}
\affiliation{Institut f\"ur Theoretische Physik, Universit\"at zu K\"oln, Z\"ulpicher Straße 77, 50937 K\"oln, Germany}

\date{\today}

 \begin{abstract}
 We provide supplemental material on the concept of free
 probability and the derivation of our
 analytical results for the spectral form factor and two-point
 function.
 \end{abstract}

\maketitle

\section{Quantum Free probability in a nutshell}
Free probability has been introduced as a mathematical framework of statistical relations
describing averages $\mathbb{E}(X_1 X_2 \dots)$ of distributed, and non-commutative $[X_i,X_j]\neq0$
elements $X_i$ of some `probability space'. We here summarize the main ideas
in less abstract terms, geared towards the analysis of time-dependent  correlation functions  $C_k\equiv \phi((A B_t)^k)$, where
$A$ and $B$ are operators in a $D$-dimensional Hilbert space,  $B_t =U^\dagger B\, U$ is time evolution, and
$\phi(X)\equiv \frac{1}{D}\, \mathbb{E}\tr(X)$.  (For notational reasons,
we here write $\mathbb{E}(\dots)\equiv \overline{(\dots)}$, the expectation value
being defined relative to a distribution of the time evolution
operators, e.g., Haar.) The architecture of
this expression motivates defining $\mathcal{O}\equiv \{A,B\}$ and
$\mathcal{U}\equiv \{U,U^\dagger\}$ as
`generators of two distinct subalgebras of a probability space'. The terminology
refers to the fact that we are in a
\emph{space} (of matrices) with an \emph{algebra} structure (matrix multiplication), where a set
of matrices such as $\mathcal{O}$ generates a subalgebra $(A,B,AB,ABA,\dots)$
containing all matrix monomials and their linear combinations.

The algebras $\mathcal{O}$ and $\mathcal{U}$ are \emph{free} relative to each
other, if the cumulants $\kappa_n(X_1,\dots,X_n)=0$ vanish whenever the argument set $\{X_i\}$ contains
at least one element from $\mathcal{O}$ and at least one from $\mathcal{U}$.
Here, $\kappa_n$ are the so-called \emph{free cumulants}, which generalize
conventional cumulants to non-commuting argument sets. While their concrete definition is somewhat tricky,
all that matters for us is that they begin linearly in their arguments, just as
conventional cumulants do. (In fact, differences to the latter start only at
fourth order, $\kappa_{\geq 4}$.) Conceptually, the freeness criterion requires
that there be no statistically connected contributions to the
expectation value involving both the random $U$'s and the fixed operators $A,B$.

Provided this condition holds, correlation functions such as $C_k$ organize
into sums
\begin{align*}
    C_k = \sum_{\pi \in \mathrm{NC}(2k)} \kappa_\pi(X)\, F_t(\pi),
\end{align*}
where $X=(A B A B \dots A B)$ is the formal word built from $2k$ letters $A$ and
$B$, and the sum extends over all non-crossing partitions $\pi$ of that word.\cite{voiculescu1992free}
For example, for $X=(A B A B)$, partitions such as $\{\{1,2\},\{3,4\}\}$ or
$\{\{1,4\},\{2,3\}\}$ are non-crossing partitions.  The object $\kappa_\pi$ is
the product of free cumulants of $A$ and $B$ associated with the blocks of
$\pi$, and $F_t(\pi)$ contains multi-trace correlators of the time evolution,
$
\mathbb{E}\!\left(\tr(U^{n_1 t})\,\tr(U^{\dagger m_1 t})\dots\right),
\sum_i n_i = \sum_j m_j = k
$. In other words, we are granted a
factorization into structural data depending on the reference operators, and
dynamical data carried by the time evolution operators.

Assuming tracelessness (or centered operators in the parlance of the field),
$\tr(A)=\tr(B)=0$, Eq.~(3) of the main text is the $k=1$ incarnation of this
hierarchy. For $k=2$ --- the `out of time order correlation function' ---
$\mathbb{E}\tr(A B_t A B_t)$ the
sum already gets significantly more complicated. It now contains the fourth free
cumulant $\kappa_4(A,B,A,B)$, dynamical factors involving the form factor at
time $2t$, $\sim \mathbb{E}|\tr(U^{2t})|^2$, as well as  correlation
$\mathbb{E}\!\left(\tr(U^{2t})\,(\tr(U^{\dagger t}))^2\right)$, etc. While this
is looking increasingly complex, the merit of freeness remains that dynamical
correlation functions continue to be determined by static cumulants depending on
$A$ and $B$, and dynamical factors depending on $U$ and $U^\dagger$ only.

\section{Path-Integral Representation}
In this section, we provide the technical details underlying the analytical results presented in the main text, generalizing the approach of Ref.~\cite{altland2025path}. We first derive the Keldysh path-integral representation of the spectral form factor and the two-point function for the minimal ancilla–environment model. After performing a color–flavor transformation, we discuss the structure of the quantum modes that govern the crossover from early- to late-time correlations.\\

\subsection{Correlation function and $\psi$-Representation}

\label{app:PsiRepresentationSingle}
We start by introducing the correlation function
\begin{align}
\label{app:correlation_function}
C_{\alpha \beta \gamma \delta}(t) 
&=
\overline{
[U^\dagger]^t_{\alpha \beta}U^t_{\gamma \delta}},
\end{align} 
with 
$\alpha = (\mu, a)$ collecting environment- and ancilla-indices, $\mu = 1, \dots, D_0$ and  $a = 0, 1$, respectively,
we express 
the spectral form factor and two-point correlation function as  
\begin{align}
K(t)= \frac{1}{D^2}\sum_{\alpha\beta\gamma\delta} C_{\alpha\beta\gamma\delta}(t) 
\delta_{\alpha\beta} \delta_{\gamma\delta}, \qquad
\qquad \expval{B_t A} = \frac{1}{D} \sum_{\alpha\beta\gamma\delta} C_{\alpha\beta\gamma\delta}(t)  A_{\delta\alpha} B_{\beta \gamma}.
\end{align}
 
Following the steps explained in 
Ref.~\cite{altland2025path}, 
we represent the latter in terms of 
the Grassmann integral, 

\begin{align}
C_{{\alpha\beta\gamma\delta}}(t) =
\int D\psi\, e^{-S[\psi]}~
\psi^-_{0\alpha}\bar{\psi}^-_{t \beta}
\psi^+_{t \gamma}\bar{\psi}^+_{0\delta}, 
\qquad     S[\psi] =  \bar \psi^+ (1- T_- U )\psi^+ + 
    \bar \psi^- (1-   U^\dagger T_+ )\psi^-,
\end{align}
where 
 $\psi^\pm= \{ \psi^\pm_{ \alpha} \}  $ is a $D$-component Grassmann vector, with $D\psi=\prod_{s=\pm}\prod_{ \alpha} d\bar \psi^s_{ \alpha} d\psi^s_{ \alpha}$, $\bar \psi \psi 
=\sum_{s=\pm}\sum_\mu \bar{\psi}^s_{ \alpha} \psi^s_{ \alpha}$,
and $s=\pm$ distinguishing the
`retarded', respectively, `advanced' time-contours $U$ and $U^\dagger$ act on,  
and
$(T_\pm \psi)_n =  \psi_{n\pm 1}$ is a time-shift operator.

\subsection{Color-Flavor Transformation}
\label{sec:CFTSingle}

We next perform the integral over Haar-random unitaries $U_0$ drawn from the Circular-Unitary Ensemble (${\rm CUE}$), 
employing an exact integral identity known as the 
Color–Flavor Transform (CFT). 
The latter
exchanges the Haar-integral for one involving matrices $Z$ in ancilla- and time-space, but structureless in the environment Hilbert-space, 
\begin{align}
    \label{eq:ColorFlavorTransform}
    \int_{\rm CUE} dU_0 ~e^{-\bar \psi^+ T_- U_g U_0 \psi^+ - \bar \psi^-  T_+ U_0^\dagger U^\dagger_g \psi^-} =\int DZ\, e^{-D_0 \tr\ln(1+Z Z^\dagger)-\bar \psi^+ 
    U_g Z_T U_g^\dagger \psi^- + \bar \psi^- Z^\dagger \psi^+}.
\end{align} 
Here `$\tr$' denotes a trace over the $2t$ ancilla space and discrete
time space, $Z=\{Z_{aa',nn'}\}$ are $2t\times 2t$ matrices in ancilla space and 
discrete time, and we have introduced 
$(Z_T)_{nn'} \equiv (T_- Z T_+)_{nn'} = Z_{(n-1)(n'-1)}$, a one-time step translated matrix. 

We then perform the Gaussian $\psi$-integral to obtain a pure $Z$-integral representation of the correlation function,
\begin{equation}
\begin{aligned}
\label{eq:BIntegralSignificance}
C_{{\alpha\beta\gamma\delta}}(t) &= 
    \int DZ ~e^{-D_0\tr\ln(1+ZZ^\dagger)} \int D \psi~ e^{-
    \bar \psi G[Z]^{-1}\psi
     } \psi^-_{0\alpha}\bar{\psi}^-_{t \beta} \psi^+_{t \gamma}\bar{\psi}^+_{0\delta}\\
&= \int DZ\, e^{-S[Z]}  \left(G^{--}_{0t,\alpha \beta} G^{++}_{t0,\gamma \delta} -  G^{-+}_{00,\alpha \delta}G^{+-}_{tt,\gamma \beta} \right),
\end{aligned}
\end{equation}where the second line follows from Wick contractions using the propagator
\begin{align}
    \label{eq:GBDefinition}
    G[B]\equiv \begin{pmatrix}
    1 &U_g Z_T U_g^\dagger \\
    -Z^\dagger &1
    \end{pmatrix}^{-1}, 
\end{align}and we have introduced the effective action for the $Z$-fields,
\begin{equation}\label{app:S[Z]-action}
    S[Z]= D_0\tr\ln(1+Z Z^\dagger)-\Tr\ln(1+U_g Z_T U_g^\dagger Z^\dagger).
\end{equation}Here, $\Tr$ denotes a trace over the $Dt$-dimensional space involving 
environment, ancilla, and discrete time.

\subsection{Pre-exponentials}
We can now express the spectral form factor and the two-point function as
\begin{equation}
\begin{aligned}
    K(t) &= \frac{1}{D^2}\biggr\langle \Tr( G_{0t}^{--}) \Tr( G_{t0}^{++}) 
    \biggr\rangle_{S[B]}, \\
    \expval{B_t A} &= \frac{1}{D}\biggr\langle \Tr( G_{0t}^{--} B G_{t0}^{++} A) - \Tr(G_{00}^{-+}A) \Tr(B G_{tt}^{+-})\biggr\rangle_{S[B]}, 
\end{aligned}
\end{equation}
where we denote $\langle \cdots\rangle_{S[Z]} = \int DZ~e^{-S[Z]} (\cdots)$.
Next, we express the various $G$ matrix elements in terms of $Z$.  At quadratic order, and up to negligible $\gamma$-dependent contributions, these are given by
\begin{equation}
    G^{rs}_{\alpha \beta, nm} = G^{rs}_{\mu \nu, ab, nm} \sim \delta_{\mu\nu}\begin{pmatrix}
    1 -  Z Z^\dagger & Z\\
    -Z^\dagger &1 -  Z^\dagger Z 
    \end{pmatrix}_{ab,nm},
\end{equation}
resulting in
\begin{equation}
\begin{aligned}
    K(t) &= \frac{1}{4}\biggr\langle \tr( Z^\dagger Z)_{0t} \tr( Z^\dagger Z)_{t0}\biggr\rangle_{S[Z]}, \\
    \expval{B_t A} &= \frac{1}{D}\biggr\langle \Tr( (Z^\dagger Z)_{0t} B (Z Z^\dagger)_{t0} A) + \Tr (Z_{00}^\dagger A) \Tr(B Z_{tt})\biggr\rangle_{S[Z]}. 
\end{aligned}
\end{equation}

\subsection{Semiclassical Expansion}

To evaluate these matrix elements, we work in the double-scaled limit 
($D_0 \rightarrow \infty$, $\gamma \rightarrow 0$), and expand the action 
Eq.~\eqref{app:S[Z]-action} to quadratic order in matrix fields,
\begin{equation}
        S_0[Z] =  -D_0\tr\left( Z^\dagger d Z + \gamma ( Z^\dagger Z  - Z^\dagger\sigma^+Z\sigma^- -Z^\dagger \sigma^-Z\sigma^+)\right).
\end{equation}
Here  $dZ = Z - T_+ Z T_-$ is the discrete derivative, and we identify the first $\gamma$-dependent contribution, $\gamma Z^\dagger Z$, as the self-energy induced damping, accounting for a finite lifetime of ancilla states due to coupling to the environment. 
The second and third contributions, $\gamma Z^\dagger (\sigma^+ Z \sigma^- + \sigma^- Z \sigma^+)$,  are vertex corrections, originating from the coherent phase cancellation for scattering events 
simultaneously occurring in the retarded and advanced propagation.

The ancilla structure of the action $S_0[Z]$ is diagonalized in the Pauli basis,
\begin{equation}
    Z = \sum_{\mu = 0}^{3} z_\mu\sigma_\mu,
\end{equation}
leading to quadratic actions for the spin singlet and triplet modes discussed in the main text.
The calculation of correlation function in this semiclassical approximation thus reduces
 to simple Gaussian integrals 
over the complex variables $z_\mu$
\begin{align} 
\langle ... \rangle_{S_0[Z]} =
\prod_{\mu=0}^3 
\int Dz_\mu \, e^{-\pi^{-1}_\mu \bar{z}_\mu z_\mu} (...), \qquad \pi_\mu (n) = \frac{1}{D} e^{-\gamma_\mu n}, \qquad \gamma_\mu = 0 ,\gamma, \gamma , 2\gamma.
\end{align}

As a result of Gaussian integrations, we obtain
\begin{equation}
    K(t) = \sum_{\mu = 0}^3 ( \pi_\mu \ast \pi_\mu  )(t) = t~ \sum_{\mu =0}^3 e^{-\gamma_\mu t},
\end{equation} 
where the condition $\mu = \nu$ is enforced by the orthogonality relation  
$\tr(\sigma_\mu \sigma_\nu) = 2 \delta_{\mu\nu}$.
Similarly, we find for operators $X=A,B$ that are traceless in the environment space, $\tr_{D_0}X = \sum_\alpha X_{\alpha a, \alpha b } = 0$,  
and thus 
\begin{equation}
    \expval{B_t A} = \sum_{\mu,\nu = 0}^3 \expval{\sigma_\mu \sigma_\nu B \sigma_\mu \sigma_\nu A} ( \pi_\mu \ast \pi_\nu  )(t),
\end{equation}
from which we obtain the universal contribution, $K(t)\expval{AB}$, only when $\mu = \nu$, 
and the non-universal correction, $\Delta(t)$, for $\mu \neq \nu$. 
The explicit relation for the two-point function follows directly from Eqs. (5) and (6), presented in the Letter.

\subsection{Semiclassical theory and Plateau}

Finally, to describe the saturation of the spectral form factor to the plateau value 
requires accounting for a second, non-perturbative saddle point. This is discussed, e.g., in Refs. \cite{Andreev1997,Altland2000} for related matrix integrals, and the adaptation of these methods to the present context leads to the spectral form factor of the ancilla-environment model
\begin{align}
K(t)  = t~ \sum_k e^{ - \gamma_k t}  - (t - D) \theta(t - D)  - \sum_{k \neq 0 } \frac{\mathcal{K}(\gamma_k)}{2\gamma_k} \left( e^{-\gamma_k |t-D|} +  e^{-\gamma_k (t+D)}\right), \qquad \qquad
    \mathcal{K}(\gamma_k) = \prod_{k'\neq 0, k}  \frac{\gamma_{k'}^2}{\gamma_{k'}^2 - \gamma_k^2},
\end{align}
accounting now also for late times $t\gtrsim t_H$, and shown in the comparison to numerical simulation in the main text.

\section*{References}
\bibliographystyle{apsrev4-2}
\bibliography{ancilla_spin_refs}